\documentclass[cits]{PoS}

\title{On the rigid string contribution to the interquark potential}

\ShortTitle{Rigid string}

\author{\speaker{Michele Caselle}, Davide Vadacchino\\
        Dipartimento di Fisica Teorica, Universit\`a di Torino\\
        and Istituto Nazionale di Fisica Nucleare, sezione di Torino,\\
        Via Pietro Giuria 1, I-10125 Torino, Italy\\
        E-mail:  \email{caselle@to.infn.it},\email{vadacchi@to.infn.it}}

\author{Marco Panero\\
       Instituto de F\'{\i}sica T\'eorica UAM/CSIC, Universidad Aut\'onoma de Madrid\\
       Calle Nicol\'as Cabrera 13-15, Cantoblanco E-28049 Madrid, Spain\\
       E-mail: \email{marco.panero@inv.uam.es}}
   
\author{Roberto Pellegrini\\
       Physics Department, Swansea University,\\
       Singleton Park, Swansea SA2 8PP, UK\\
       E-mail: \email{ropelleg@to.infn.it}}
       
\abstract{Owing to Lorentz invariance, the leading terms in the effective action describing the low-energy dynamics of flux tubes in a confining gauge theory are universal. Besides the terms corresponding to the Nambu-Goto action, Lorentz invariance also allows terms describing a ``rigid string'', proposed long ago by Kleinert and Polyakov as a model for confinement in a dual superconductor scenario. In this contribution we discuss some of the non-trivial properties of this type of string and evaluate analytically the corrections to the interquark potential induced by extrinsic curvature terms appearing in the rigid string action.
\vspace{1cm}

\begin{flushright}
IFT-UAM/CSIC-14-108
\end{flushright}
}

\FullConference{The 32nd International Symposium on Lattice Field Theory,\\
		23-28 June, 2014\\
		Columbia University New York, NY}

\newcommand{\Z}{\mathbb{Z}}

\newcommand{\SU}{\mathrm{SU}}
\newcommand{\U}{\mathrm{U}}

\newcommand{\SNG}{S_{\mbox{\tiny{NG}}}}
\newcommand{\Scl}{S_{\mbox{\tiny{cl}}}}
\newcommand{\SPol}{S_{\mbox{\tiny{Pol}}}}
\newcommand{\Sb}{S_{\mbox{\tiny{b}}}}
\newcommand{\VNG}{V_{\mbox{\tiny{NG}}}}
\newcommand{\Vb}{V_{\mbox{\tiny{b}}}}
\newcommand{\Vext}{V_{\mbox{\tiny{ext}}}}

\newcommand{\re}{{\rm{Re}}}

\newcommand{\eq}{\begin{equation}}
\newcommand{\en}{\end{equation}}
\newcommand{\eqar}{\begin{eqnarray}}
\newcommand{\enar}{\end{eqnarray}}

\begin{document}
\section{Introduction}
\label{sect:intro}

As it is well-known, a confining flux tube in Yang-Mills theory can be modeled in terms of a fluctuating bosonic string, governed by a suitable effective action. When expanded around the low-energy limit, such effective action can be decomposed into the sum of different terms, with coefficients which, a priori, depend on the Yang-Mills theory under consideration. However, in a series of recent works~\cite{Luscher:2004ib, Aharony:2009gg, Meyer:2006qx, Aharony:2011gb, Gliozzi:2011hj, Gliozzi:2012cx, Meineri:2013ew, Aharony:2013ipa} it was demonstrated that the first few terms in the effective action are actually \emph{universal}. This universality follows directly from the symmetry
constraints one must impose on the action, and makes this effective theory much more predictive than other effective models in particle physics.

The easiest way to understand how these constraints arise is by going to the so-called ``physical gauge'', whereby the two worldsheet coordinates are identified with the longitudinal degrees of freedom of the string: $\xi^0=X^0$, $\xi^1=X^1$. Then, the string action becomes a function of the remaining $(D-2)$ degrees of freedom, which correspond to the transverse displacements of the string worldsheet, $X^i$, with $i=2, \dots , (D-1)$. 
In the low-energy limit, the action can be expanded into a sum of terms with an increasing number of derivatives of the transverse degrees of freedom of the string: the first few terms in this expansion read
\eq
S=\Scl+\frac\sigma2\int d^2\xi\left[\partial_\alpha X_i\cdot\partial^\alpha X^i+
c_2(\partial_\alpha X_i \cdot\partial^\alpha X^i)^2
+c_3(\partial_\alpha X_i \cdot\partial_\beta X^i)^2+\dots\right]\,,
\label{action}
\en
where the classical action $\Scl$ includes a term corresponding to the minimal string worldsheet area (and a possible perimeter term), the second term represents a massless free field in two dimensions~\cite{Luscher:1980fr}, while the other terms correspond to higher-order interactions of the $X_i$ fields. 

The $c_i$ coefficients must satisfy a set of constraints, to enforce Lorentz invariance. Indeed, even though the Lorentz invariance of the original theory is broken by the classical string configuration around which one is expanding, the effective action should still be consistent with this symmetry, by means of a non-linear realization in terms of the $X_i$ fields~\cite{Meyer:2006qx, Aharony:2011gb, Gliozzi:2011hj, Gliozzi:2012cx, Meineri:2013ew}. These constraints lead to recursive relations among the expansion coefficients, and dramatically reduce the number of free parameters of the effective theory. One can show that the terms with only first derivatives are consistent with the Nambu-Goto action, to all orders in the derivative expansion~\cite{Aharony:2010cx}. For the next-to-leading-order terms (i.e. those beyond the Nambu-Goto action) one can find~\cite{Gliozzi:2012cx, Meineri:2013ew,Aharony:2013ipa} that the first correction allowed by Lorentz invariance is (the gauge-fixed version of) a term proportional to the square of the extrinsic curvature $K$,
\begin{equation}
S_{2,K}=\alpha\int d^2\xi \sqrt{g} K^2\,. \label{ext}
\end{equation}
Classically, this term could be eliminated by a suitable redefinition of the $X_i$ fields, but it can be shown that this redefinition is anomalous~\cite{Caselle:2014eka}, so that the term in eq.~(\ref{ext}) can give a contribution at one loop (which we shall discuss in
detail below). In particular, this contribution is proportional to the parameter $\alpha$ which, at this order, is the only other free parameter of the effective string description besides the string tension $\sigma$. Finally, Lorentz invariance also implies that in three dimensions the leading correction to the Nambu-Goto string appears only at a high order.\footnote{In particular, for the ground-state charge-anticharge potential 
$V(R)$ (with $R$ denoting the distance between the static sources, taken to be large) the first new correction appears at $O(R^{-7})$.} This allows one to carry out stringent tests of the effective string picture and, at least in principle, to extract the value of $\alpha$. 
Indeed, in the past few years it has been possible to observe tiny deviations from 
the pure Nambu-Goto behavior in high-precision lattice studies~\cite{Athenodorou:2010cs, Caselle:2007yc, Caselle:2010pf, Brandt:2010bw}. These deviations were observed both 
in $\SU(N)$ Yang-Mills theories~\cite{Athenodorou:2010cs, Brandt:2010bw} and in the ground state potential of the $\Z_2$ lattice gauge theory in three dimensions (3D), where deviations were found for worldsheets with torus~\cite{Caselle:2007yc} and cylinder topology~\cite{Caselle:2010pf}. These deviations are compatible with a non-zero value of $\alpha$, but they are so small, that until now it was not possible to extract a precise value of $\alpha$ and test the above picture. On the other hand, in the 3D $\U(1)$ theory, macroscopic deviations from the expectations described above can be observed for a wide range of distances and of $\beta=1/(a e^2)$ (where $a$ denotes the lattice spacing and $e$ is the coupling; note that in 3D the latter has dimension $1/2$)~\cite{Caselle:2014eka, Vadacchino:2013baa}, thus suggesting a value of $\alpha$ much larger than in the gauge theories mentioned above. These results are the first confirmation of an earlier theoretical prediction by Polyakov, who suggested that the effective string action for this model could include an extrinsic-curvature term~\cite{Polyakov:1996nc}. In this respect, the 3D $\U(1)$ model represents a unique tool to explore the effects of the extrinsic curvature terms. In this contribution we review the derivation, using the zeta-function regularization, of the extrinsic curvature contribution to the interquark potential and discuss the non-trivial scaling behavior of $\alpha/\sigma$ in the continuum limit (which explains why the 3D U(1) model  behaves differently from all the other LGTs studied up to now). In a companion contribution~\cite{Vadacchino_Lattice_2014} we compare these predictions with a set of high precision numerical data and address the $\beta$ dependence of the ratio $\alpha/\sigma$.

\section{$\U(1)$ gauge theory in three spacetime dimensions}
\label{sect:U1}

The lattice Wilson action of the $\U(1)$ gauge theory in an isotropic cubic lattice $\Lambda$ of spacing $a$ in three spacetime dimensions is defined as
\eq
\label{Wilson_action}
\beta \sum_{x \in \Lambda} \sum_{1 \le \mu < \nu \le 3} \left[1 - \re \, U_{\mu\nu}(x)\right], \;\;\; \mbox{with} \;\;\; \beta=\frac{1}{a e^2},
\en
where 
\eq
\label{plaquette}
U_{\mu\nu}(x) = U_\mu(x) U_\nu(x+a\hat\mu) U^\star_\mu(x+a\hat\nu) U^\star_\nu(x), \;\;\; \mbox{with} \;\;\; U_\mu(x) = \exp \left[ i a A_\mu \left( x + a \hat{\mu}/2 \right) \right].
\en
Remarkably, this model can be studied analytically 
in the semi-classical approximation~\cite{Polyakov:1976fu, Gopfert:1981er}:
this leads to the result that the model is confining for all values of $\beta$, and that in the $ \beta\gg 1$ limit it reduces to a theory of free massive scalars. Moreover, in this limit, an explicit expression for the lightest glueball and a lower bound for the string tension can be explicitly derived:
\begin{equation}
m_0 a = c_0\sqrt{8 \pi^2 \beta}e^{-\pi^2\beta v(0)},\;\;\; \sigma a^2 \geq
\frac{c_{\sigma}}{\sqrt{2\pi^2\beta}} e^{-\pi^2\beta v(0)},
\end{equation}
where $c_0=1$ and $c_{\sigma}=8$ in the semi-classical approximation. 

Lattice Monte Carlo computations~\cite{Caselle:2014eka, Loan:2002ej} show that the string tension saturates this bound and that, while both constants are affected by the semi-classical approximation and change their values in the continuum limit, they are always positive, so the model is confining at any $\beta$, even beyond the semi-classical approximation. In contrast to usual confining gauge theories in four dimensions, in which the $m_0/\sqrt{\sigma}$ ratio is fixed, using this lattice model at finite $a$ one has
\begin{equation}
\label{ratio_m0}
\frac{m_0}{\sqrt{\sigma}} = \frac{2 c_0}{\sqrt{c_\sigma}} 
(2\pi^2\beta)^{3/4} e^{-\pi^2 v(0)\beta/2}\,
\end{equation}
so by tuning $\beta$ it is possible to set the $m_0/\sqrt{\sigma}$ ratio in the lattice model to any desired value.

\subsection{A rigid-string description for the $\U(1)$ model}

Another interesting feature of this model is that it allows, at least qualitatively, an explicit string description. An argument for this string description was put forward by Polyakov in ref.~\cite{Polyakov:1996nc}: he combined the Nambu-Goto action with an extrinsic curvature term into a so-called `rigid string'' action. Polyakov also discussed how the couplings in this effective action depend on the charge and on the mass gap of the underlying gauge theory 
(see ref.~\cite[eq.~(17)]{Polyakov:1996nc}). This string action reads
\eq
\SPol= c_1 e^2 m_0 \int d^2\xi \sqrt{g}~ + c_2 \frac{e^2}{m_0} \int d^2\xi \sqrt{g} K^2 \, ,
 \label{Polyakov}
 \en
where $c_1$ and $c_2$ are two undetermined constants. Identifying the latter with $\sigma$ and $\alpha$ one obtains $\sqrt{\sigma/\alpha} \sim m_0$. This result will play a crucial r\^ole in our discussion.

\section{Effective string action and extrinsic curvature}
\label{sect:string}

In addition to the Nambu-Goto and extrinsic curvature terms, the most general form of the rigid string action can include a boundary term $\Sb$
\eq
S = \SNG + S_{2,K} + \Sb \, .
\label{action2}
\en
Like the other terms, also $\Sb$ is constrained by Poincar\'e symmetry: when the boundary represents a Polyakov line at $\xi_1=0$, in the $\xi_0$ direction and with Dirichlet boundary conditions $X_i(\xi_0,0)=0$, in the physical gauge one can expand $\Sb$ as
\eq
\Sb=\int d\xi_0 \left[b_1\partial_1 X_i\cdot \partial_1 X^i+b_2\partial_1\partial_0 X_i\cdot \partial_1\partial_0 X^i+
\dots\right]\,.
\label{bounda}
\en
In 3D, there is only one transverse degree of freedom, described by a single bosonic field $X(\xi_1,\xi_2)$. Keeping only terms up to four derivatives, one finds: 
\begin{eqnarray}
&& \SNG \simeq 
\Scl+\frac\sigma2\int d^2\xi\left[\partial_\alpha X\cdot\partial^\alpha X-
\frac14(\partial_\alpha X \cdot\partial^\alpha X)^2\right]\,,
\label{action3} \\
&& S_{2,K} \simeq \alpha\int d^2\xi (\Delta X)^2,
\label{action4} \\
&& \Sb \simeq b_2\int d\xi_0 \left[\partial_1\partial_0 X\cdot \partial_1\partial_0 X
\right]\,. \label{bounda2}
\end{eqnarray}
Thus we are left with three free parameters ($\sigma$, $\alpha$ and $b_2$) which, in principle, could be fitted to numerical lattice results.

The string described by this action is known in the literature as ``rigid string''~\cite{Polyakov:1986cs, Kleinert:1986bk}. Its contribution to the interquark potential was evaluated in refs.~\cite{Braaten:1986bz, German:1989vk, Nesterenko:1997ku}. We report here the results relevant for the case in which the potential is extracted from  Polyakov loop correlators in the low-temperature limit, with fixed boundary conditions along the spatial direction of separation between the loops, $X(t,0)=X(t,R)=0$ and for $R \ll N_t$.
The analysis of refs.~\cite{Caselle:2014eka, Nesterenko:1997ku} shows that the interquark potential can be written as the sum of six terms:
\begin{equation}
V(R)= \sigma R + \VNG(R)+\VNG'(R,\sigma) + \Vext(R,m)+\Vext'(R,m,\sigma) + \Vb\,,
\end{equation}
with:
\begin{eqnarray}
&& \VNG(R)
=-\frac{\pi}{24 R}\,, \\
&& \VNG'(R,\sigma)
=-\left(\frac{\pi}{24}\right)^2\frac{1}{2\sigma R^3}\,, \\
&& \Vext(R,m)
=-\frac{m}{2\pi}\sum\limits_{n=1}^{
\infty} \frac{K_1 \left(2n m R\right) }{n}\,,\\
&& \Vext'(R,m,\sigma)
=\left(\frac{\pi}{24}\right)^2\frac{21}{20 m\sigma R^4}\,,\\
&&
\Vb(R,b,\sigma)=-b_2\frac{\pi^3}{60 R^4}\,,
\end{eqnarray}
where $m\equiv \frac{\sigma}{2\alpha}$ encodes the magnitude of the extrinsic curvature term,  $K_\alpha(z)$ is a modified Bessel function of the second kind and we have separated in $\VNG$ and $\Vext$ the $\sigma$-independent Gaussian part from the corrections due to the quartic term in the Nambu-Goto action, denoted as $\VNG'$ and $\Vext'$.

Some remarks about this result are in order.
\begin{itemize}
\item The Gaussian part of the extrinsic curvature potential $\Vext(R,m)$ 
 coincides with the partition function of a massive perturbation of the $c=1$ free bosonic theory.
Indeed, the quantity 
\eq
c_0(mR) = -\frac{24 R}{\pi} \Vext(R,m)
\en
is nothing but the ground-state scaling function that was introduced in ref.~\cite{Klassen:1990dx} to describe such perturbation. $c_0(mR)$ is monotonically decreasing, interpolates between $1$ (when $mR=0$) and $0$ (in the $mR \to \infty$ limit).
It is interesting to note the analogy with the Nambu-Goto model: while the latter can be interpreted as an irrelevant \emph{massless} perturbation of the two-dimensional bosonic conformal field theory (CFT) with central charge $c=1$~\cite{Dubovsky:2012sh, Caselle:2013dra}, the rigid string corresponds to a relevant \emph{massive} perturbation of the same CFT. 
\item The presence of a massive degree of freedom on the worldsheet of the confining string might explain the deviations from the Nambu-Goto model observed in lattice studies of $\SU(N)$ Yang-Mills theories~\cite{Dubovsky:2013gi, Dubovsky:2014fma}. Our results provide a realization of this scenario in the $\U(1)$ model in 3D.
\item For $mR \to 0$, one recovers a free bosonic theory, with a second ``L\"uscher'' term, besides the one from $\VNG(R)$. Therefore at small $m$ ($m < \pi \sqrt{\sigma}$), the extrinsic curvature term should have a dramatic impact onto the finite-size corrections to the interquark potential. If $m \propto m_0$, as suggested by Polyakov, see eq.~(\ref{Polyakov}), then eq.~(\ref{ratio_m0}) implies that the contribution of the extrinsic curvature grows for increasing $\beta$, and becomes dominant for $a \to 0$.
\item
At large $R$, the behavior of $\Vext(R,m)$ is dominated by the lowest-index Bessel function appearing in the sum:
\eq
\Vext(R,m) \simeq -\sqrt{\frac{m}{16\pi R}} e^{-2mR} \qquad \mbox{for} \;\; R \gg \frac{1}{m} ,
\en
resulting in an exponential decrease of $\Vext(R,m)$. This behavior is characteristic of a massive perturbation of a 2D CFT, and is consistent with results about how the extrinsic curvature coupling for the rigid string changes under renormalization-group transformations~\cite{Polyakov:1986cs, Kleinert:1986bk}. Thus, taking the continuum limit at fixed $m$ the contribution of $\Vext(R,m)$ becomes negligible, but in the continuum limit \emph{at fixed} $mR$ its contribution can be important. 
\end{itemize}

\section{Concluding remarks}
\label{sect:conclusions}
In this contribution, we discussed the main corrections to the interquark potential due to the rigid string action, in particular those related to the extrinsic curvature term. These corrections are a function of the ratio $m=\sigma/(2\alpha)$ and, in the continuum limit, their relative magnitude with respect to the Nambu-Goto corrections depends on the $m/\sqrt{\sigma}$ ratio. For most gauge theories of interest, this ratio is constant and the extrinsic curvature term gives only a small correction to the interquark potential. 
However, for the 3D $\U(1)$ model, Polyakov's derivation suggests to relate the parameter $m$, which controls the rigid-string contribution, to the mass of the lightest glueball $m_0$ and thus, since in the $\U(1)$ model $m_0/\sqrt{\sigma}\sim m/\sqrt{\sigma}$ decreases exponentially with $\beta$, we expect the continuum limit of the model to be dominated by rigid-string behavior. Indeed our recent Monte Carlo study~\cite{Caselle:2014eka}, discussed in a companion contribution to these proceedings~\cite{Vadacchino_Lattice_2014}, shows that in this model, as $\beta$ increases towards the continuum limit, the interquark potential shows strong deviations from a Nambu-Goto effective string model, and that these deviations are described well by the one-loop contribution of an extrinsic curvature term in the effective string action.

Clearly, a rigid string (i.e. a string dominated by the extrinsic curvature term) is very different from the Nambu-Goto one: leads to different properties not only for the interquark potential but also for the string thickness and for the deconfinement temperature. In this sense, it is really ``a different kind of string''~\cite{Caselle:2014eka}. The 3D $\U(1)$ lattice model turns out to be a perfect laboratory to study the cross-over from a Nambu-Goto string at small $\beta$ to a rigid string at large $\beta$.

Our findings may have important implications for other confining theories, too, including some of interest for elementary particle physics (like the $\SU(3)$ theory in four spacetime dimensions) or for condensed matter physics (like the 3D Ising model). In these cases the extrinsic curvature term only gives subleading corrections, which however could explain the small deviations from Nambu-Goto string behavior observed in recent simulations~\cite{Athenodorou:2010cs, Caselle:2007yc, Caselle:2010pf}.

{\bf Acknowledgments} This work is supported by the Spanish MINECO (grant FPA2012-31686 and ``Centro de Excelencia Severo Ochoa'' programme grant SEV-2012-0249).

\end{document}